\def\fr#1#2{\hbox{${#1\over #2}$}}
\def\subsub#1{\paragraph*{\bf #1}}   
\def\lra{\leftrightarrow}           
                  \def\pd{\partial}
\def\ort{\perp}                     \def\tgr{GR$_{\parallel}$}
\def\mb#1{\hbox{\boldmath$#1$}}     
\def\bT{\bar T}                     
\def\bcH{{\bar{\cal H}}}            \def\bphi{\bar\phi}

\def\m{\mu}             \def\n{\nu}              \def\k{\kappa}
          \def\g{\gamma}           \def\d{\delta}
          \def\s{\sigma}           \def\t{\tau}
\def\a{\alpha}          \def\b{\beta}            \def\th{\theta}
\def\vphi{\varphi}      \def\ve{\varepsilon}
\def\r{\rho}                       \def\p{\pi}
\def\l{\lambda}         \def\o{\omega}           \def\O{\Omega}
\def\bn{\bar n}         \def\bi{{\bar\imath}}    \def\bk{{\bar k}}
       \def\bm{{\bar m}}        \def\bn{{\bar n}}
   
\def\cL{{\cal L}}       \def\cH{{\cal H}}        \def\cP{{\cal P}}
\def\cM{{\cal M}}       \def\cO{{\cal O}}        \def\cE{{\cal E}}
\def\hp{{\hat\pi}}      \def\hcO{\hat\cO}        \def\hcH{\hat\cH}
           \def\tP{\tilde P}
\def\tM{\tilde M}       \def\tH{\tilde H}

\def\TA{\stackrel{A}{T}}
\def\one{{(1)}}         \def\two{{(2)}}
\def\osl{\symbol{'034}}
\documentstyle[preprint,aps,eqsecnum]{revtex}
\tighten
\def\nn{\nonumber}
\def\be{\begin{equation}}             \def\ee{\end{equation}}
\def\ba{\begin{array}{ll}}            \def\ea{\end{array}}
\def\bea{\begin{eqnarray} }           \def\eea{\end{eqnarray} }
\def\lab#1{\label{eq:#1}}             \def\eq#1{(\ref{eq:#1})}
\def\bsubeq{\begin{mathletters}}      \def\esubeq{\end{mathletters}}
\def\bitem{\begin{itemize}}           \def\eitem{\end{itemize}}

\begin{document}
\title{Conservation laws in the teleparallel theory of gravity}
\author{M. Blagojevi\'c$^{1,2}$ and M. Vasili\'c$^{1,}$\thanks{
        Email addresses: mb@phy.bg.ac.yu and mvasilic@phy.bg.ac.yu}}
\address{$^{1}$Institute of Physics, 11001 Belgrade, P. O. Box 57, Yugoslavia\\
         $^{2}$PINT, 6001 Koper, P. O. Box 327, Slovenia}
%\date{\today}
\maketitle
\begin{abstract}
We study the conservation laws associated with the asymptotic
Poincar\'e symmetry of spacetime in the general teleparallel theory
of gravity. Demanding that the canonical Poincar\'e generators have
well defined functional derivatives in a properly defined phase
space, we obtain the improved form of the generators, containing
certain surface terms. These terms are shown to represent the values
of the related conserved charges: energy-momentum and angular
momentum.
\end{abstract}
%pacs{PACS numbers: 04.50.+h, 04.20.Fy}

\section{Introduction}

A field theory is defined by both the field equations and boundary
conditions. In contrast to the usual flat-space field theories, the
boundary conditions in gravitational theories define the asymptotic
structure of spacetime. The concept of asymptotic or boundary symmetry
is of fundamental importance for understanding the conservation laws in
gravity; it is defined by the gauge transformations that leave a chosen
set of boundary conditions invariant. The asymptotic symmetry has a
very clear dynamical interpretation: the symmetry of the action breaks
down to the symmetry of boundary conditions, which plays the role of
the physical symmetry and defines the corresponding conservation laws.
A consistent picture of the gravitational energy and other conserved
quantities in general relativity (GR) emerged only after the role of
boundary conditions and their symmetries had been fully recognized
\cite{1,2,3}.

The Poincar\'e gauge theory \cite{4,5,6} (PGT) is a natural extension of the
gauge principle to spacetime symmetries, and represents a viable
alternative to general relativity (GR) (for more general attempts see
Ref. \cite{7}). A particularly interesting limit of PGT is given by the
teleparallel geometry \cite{8,9,10,11}. The teleparallel description of
gravity has been a promising alternative to GR until the work of
Kopczy\'nski \cite{12}, who found a hidden gauge symmetry which prevents the
torsion from being completely determined by the field equations, and
concluded that the theory is inconsistent. Possible consequences of
this conclusion have been further discussed by M\"uller-Hoisson and
Nitsch \cite{x1}. Nester \cite{13} improved the arguments by showing that the
unpredictable behavior of torsion occurs only for some very special
solutions (see also Refs. \cite{14}).

The canonical analysis of the teleparallel formulation of GR
\cite{15} (see also Ref. \cite{16}) is an important step towards
clarifying the gauge structure of the teleparallel theory. In this
case, the presence of non-dynamical torsion components is shown to
be not a sign of an inconsistency, but a consequence of the
constraint structure of the theory. The undetermined torsion
components appear as a consequence of extra gauge symmetries  (with
respect to which the torsion tensor is not a covariant object). In
the general teleparallel theory, the influence of extra gauge
symmetries on the existence of a consistent coupling with matter is
presently not completely clear. We shall assume that matter coupling
respects all extra gauge symmetries of the gravitational sector, if
they exist.

Hecht et. al. \cite{x2} investigated the initial value problem of
the teleparallel form of GR, and concluded that it is well defined
if the undetermined velocities are dropped out from the set of
dynamical velocities. The problem has not been analyzed for more
general teleparallel theories, but the results of Hecht et. al.
\cite{x3}, related to $T^2$ theories in $U_4$, are encouraging.
Among various successful applications of the teleparallel approach,
one should mention a pure tensorial proof of the positivity of
energy in GR \cite{x4}, a transparent introduction of Ashtekar's
complex variables \cite{x5}, and a formulation of a five-dimensional
teleparallel equivalent of the Kaluza-Klein theory \cite{x6}.
Although quantum properties of PGT are in general not so attractive
\cite{x7,x8}, the related behavior in the specific case of the
teleparallel theory might be better \cite{x9,x10}, and should be
further explored.

At the classical level, further progress has been made by carrying
out an explicit construction of the generators of all gauge
symmetries of the general teleparallel theory \cite{17}. In the
present paper, we continue this investigation by studying the
important relation between asymptotic symmetries and conserved
charges for isolated gravitating systems.  Assuming that the
symmetry in the assymptotic region is given by the global Poincar\'e
symmetry, we shall use the Regge-Teitelboim approach to find out the
form of the improved canonical generators \cite{3,18}. The method is
based on the fact that the canonical generators act on dynamical
variables via Poisson brackets, which implies that they should have
well defined functional derivatives. The global Poincar\'e
generators do not satisfy this requirement unless we redefine them
by adding certain surface terms, which are shown to represent the
conserved values of the energy-momentum and angular momentum.

Basic gravitational variables in PGT are the tetrad field $b^k{_\m}$
and the Lorentz connection $A^{ij}{_\m}$, and their field strengths
are geometrically identified with the torsion  and the curvature,
respectively: $T^k{}_{\m\n}=\pd_\m
b^k{_\n}+A^k{}_{s\m}b^s{_\n}-(\m\lra\n)$, $R^{ij}{}_{\m\n}=\pd_\m
A^{ij}{_\n}+A^i{}_{s\m}A^{sj}{_\n}-(\m\lra\n)$. General geometric
structure of PGT is described by the Riemann-Cartan space $U_4$,
possessing metric (or tetrad) and metric compatible connection. The
teleparallel or Weitzenb\"ock geometry $T_4$ can be formulated as a
special limit of PGT, characterized by a metric compatible
connection possessing non-vanishing torsion, while the curvature is
restricted to vanish (see also \cite{y1}):
\be
R^{ij}{}_{\m\n}(A)=0 \, .                                   \lab{1.1}
\ee
Teleparallel theories describe the dynamical content of spacetime by
a class of Lagrangians  quadratic in the torsion:
\bea
&&\cL = b\bigl(\cL_T + \cL_M\bigr)
       +\l_{ij}{}^{\m\n}R^{ij}{}_{\m\n}\, ,\nn\\
&&\cL_T=a\bigl(AT_{ijk}T^{ijk}+BT_{ijk}T^{jik}+CT_{k}T^{k}\bigr)
       \equiv \b_{ijk}(T)T^{ijk} \, .                       \lab{1.2}
\eea
Here, $\cL_M$ is the matter Lagrangian, the Lagrange multipliers
$\l_{ij}{}^{\m\n}$ ensure the teleparallelism condition \eq{1.1},
$A,B$ and $C$ are free parameters \cite{f1}, $a=1/2\k$ ($\k$
is Einstein's gravitational constant), and $T_k=T^m{}_{mk}$. The
gravitational field equations, obtained by varying $\cL$ with
respect to $b^i{_\m},A^{ij}{_\m}$ and $\l_{ij}{}^{\m\n}$, have the
form: \bsubeq \lab{1.3}
\bea
&&4\nabla_\r\bigl(b\b_i{^{\m\r}}\bigr)
   -4b\b^{nm\m}T_{nmi}+h_i{^\m}b\cL_T =\t^\m{_i}\, ,     \lab{1.3a}\\
&&4\nabla_\r\l_{ij}{^{\m\r}}
  -8b\b_{[ij]}{^\m}=\s^\m{}_{ij}\, ,                     \lab{1.3b}\\
&&R^{ij}{}_{\m\n}=0 \, ,                                 \lab{1.3c}
\eea
\esubeq
where $\t^\m{_i}$ and $\s^\m{}_{ij}$ are the energy-momentum
and spin currents of matter fields, respectively. If the
gravitational sector of the theory possesses extra gauge symmetries,
the matter coupling is assumed to respect them. The physical
interpretation of the teleparallel theories is based on the
observation that there exists a one-parameter family of the
teleparallel Lagrangians \eq{1.2}, defined by the condition
$i)~2A+B+C=0,~C=-1,$ which represents a viable gravitational theory
for macroscopic matter, observationally indistinguishable from GR
\cite{9,10,11}. For the parameter value $ii)~B=1/2$, the gravitational
part of (1.2)  coincides, modulo a four-divergence, with the
Hilbert-Einstein Lagrangian, and yields the teleparallel form of GR,
\tgr.

The layout of the paper is as follows. In Sec. II, we construct the
generators of the asymptotic (global) Poincar\'e symmetry from the
corresponding local expressions \cite{17}, and introduce an
appropriate asymptotic structure of the phase space in which these
generators act. Then, in Sec. III, we impose the requirement that
the asymptotic Poincar\'e generators have well defined functional
derivatives. This leads to the appearance of certain surface terms
in the generators, which define the values of the related charges:
energy-momentum and angular momentum of the gravitating system. In
the next section, we discuss the conservation laws of these charges,
which are associated with the asymptotic symmetry of spacetime. In
Sec. V, we transform our expressions for the conserved charges into
the Lagrangian form, and compare them with the known GR results.
Finally, Sec. VI is devoted to concluding remarks, while some
technical details are presented in the Appendices.

Our conventions are the same as in Refs. \cite{15,16,17,18}: the
Latin indices refer to the local Lorentz frame, the Greek indices
refer to the coordinate frame; the first letters of both alphabets
$(a,b,c,\dots;\a,\b,\g,\dots)$ run over $1,2,3$, and the middle
alphabet letters $(i,j,k,\dots;\m,\n,\l,\dots)$ run over $0,1,2,3$;
$\eta_{ij}=(+,-,-,-)$, and $\ve^{ijkl}$ is completely antisymmetric
symbol normalized by $\ve^{0123}=+1$.

\section{Asymptotic symmetry of spacetime}

We begin our considerations by discussing the asymptotic structure
of spacetime and the related form of the symmetry generators in the
teleparallel theory. Our attention will be limited to gravitating
systems which are characterized by the global Poincar\'e symmetry at
large distances.

\subsection{Poincar\'e invariance in the asymptotic region}

The generators of local Poincar\'e transformations in the
teleparallel theory of gravity without matter are constructed in
Ref. \cite{17} (Appendix A). The global Poincar\'e transformations
of fields and momenta are obtained from the corresponding local
transformations, given in Eqs. (2.3) and (5.2) of Ref. \cite{17}, by
the replacements
\be
\o^{ij}(x)\to\ve^{ij}\, ,\qquad
\xi^\m(x)\to \ve^\m{_\n}x^\n +\ve^\m\, ,                    \lab{2.1}
\ee
where $\ve^{ij},\ve^\m$ are constants, and
$\ve^\m{_\n}=\d_i^\m\eta_{j\n}\ve^{ij}$. We use the usual convention
according to which indices of quantities related to the asymptotic
spacetime are raised and lowered by the Minkowski metric, while the
transition between local Lorentz and coordinate basis is realized
with the help of the Kronecker symbol. In an analogous manner, the
global Poincar\'e generators are obtained from the corresponding
local expressions \eq{A4}:
\bsubeq\lab{2.2}
\bea
G=-\ve^\m P_\m +\fr{1}{2}\ve^{ij}M_{ij} \, ,               \lab{2.2a}
\eea
where
\bea
&&P_\m =\int d^3x\cP_\m\, ,\qquad
   M_{\m\n}=\int d^3 x\cM_{\m\n}\, ,\nn\\
&&\cM_{\a\b}=-S_{\a\b}+x_\a\cP_\b-x_\b\cP_\a \nn\\
&&\cM_{0\b}=-S_{0\b}+x_0\cP_\b-x_\b\cP_0
   +b^k{_\b}\p_k{^0}+\fr{1}{2}A^{ij}{_\b}\p_{ij}{^0}
   -\fr{1}{2}\l_{ij}{}^{0\g}\p^{ij}{}_{\b\g}  \, .         \lab{2.2b}
\eea
\esubeq
The quantities $\cP_\m$ and $S_{\m\n}=\d^i_\m\d^j_\n S_{ij}$ are
defined in Appendix A.

\subsection{Boundary conditions}

We shall be concerned here with {\it finite\/} gravitational
sources, characterized by matter fields which decrease sufficiently
fast at large distances, so that their contribution to surface
integrals vanishes. Consequently, the matter field contribution to
the symmetry generators can be effectively ignored. In that case, we
can assume that spacetime is {\it asymptotically flat\/}, i.e. that
the following two conditions are fulfilled \cite{19,20}:
\bitem
\item[$a)$] there exists a coordinate system in which the metric
$g_{\m\n}$ of spacetime becomes Minkowskian at large distances:
$g_{\m\n}=\eta_{\m\n}+\cO_1$, where $\cO_n=\cO(r^{-n})$ denotes a
term which decreases like $r^{-n}$ or faster for large $r$, and
$r^2=(x^1)^2+(x^2)^2+(x^3)^2$; \par
\item[$b)$] the Lorentz field strength defines the absolute
parallelism for large $r$:
$R^{ij}{}_{\m\n}=\cO_{2+\a}$, with $\a>0$.
\eitem

The first conditions is self evident, and the second one is
trivially satisfied in the teleparallel theory, where
$R^{ij}{}_{\m\n}(A)=0$. The vanishing of the curvature means that
the Lorentz gauge potential $A^{ij}{_\m}$ is a pure gauge, hence it
can be transformed to zero by a suitable local Lorentz
transformation. Therefore, we can adopt the following asymptotic
behavior of the translational and Lorentz gauge fields:
\bsubeq\lab{2.3}
\be
b^i{_\m}=\d^i_\m+\cO_1 \, ,\qquad A^{ij}{_\m}=\hcO\, ,     \lab{2.3a}
\ee
where $\hcO$ denotes a term with an arbitrarily fast asymptotic
decrease.

There is one more Lagrangian variable in the teleparallel theory,
the Lagrange multiplier $\l_{ij}{}^{\m\n}$, which is not directly
related to the above geometric conditions $a)$ and $b)$. Its
asymptotic behavior can be determined with the help of the second
field equation, leading to
\be
\l_{ij}{}^{\m\n}=\hbox{const.}+\cO_1\,.                  \lab{2.3b}
\ee
\esubeq

The vacuum values of the fields, $b^i{_\m}=\d^i_\m, A^{ij}{_\m}=0$
and $\l_{ij}{}^{\m\n}=\hbox{const.}$, are taken to be invariant
under the action of the global Poincar\'e group. Demanding that the
asymptotic conditions \eq{2.3} be invariant under the global
Poincar\'e transformations, we obtain the following conditions on
the field derivatives:
\bea
&&b^k{}_{\m,\n}=\cO_2\, ,\qquad b^k{}_{\m,\n\r}=\cO_3\, ,\nn\\
&&\l_{ij}{}^{\m\n}{}_{,\r}=\cO_2\, ,\qquad
   \l_{ij}{}^{\m\n}{}_{,\r\l}=\cO_3 \, .                   \lab{2.4}
\eea
The above relations impose serious restrictions on the gravitational
field in the asymptotic region, and define an {\it isolated\/}
gravitational system (characterized, in particular, by the absence
of gravitational waves).

The asymptotic behavior \eq{2.3} and \eq{2.4} is compatible not only
with global Poincar\'e transformations, but also with a restricted
set of local Poincar\'e transformations, whose gauge parameters
decrease sufficiently fast for large $r$.

In addition to the asymptotic conditions \eq{2.3} and \eq{2.4}, we
shall adopt the principle that {\it all the expressions that vanish
on shell have an arbitrarily fast asymptotic decrease\/}, as no
solution of the equations of motion is thereby lost. In particular,
the constraints of the theory are assumed to decrease arbitrarily
fast, and consequently, all volume integrals defining the Poincar\'e
generators \eq{2.2} are convergent.

The asymptotic behavior of momentum variables is determined using
the asymptotics for the fields and the relation
$\p-\pd\cL/\pd\dot\vphi=\hcO$, in accordance with the above
principle. Thus, we find
\bea
&&\p_i{^0},\p_{ij}{^0},\p^{ij}{}_{\m\n}=\hcO\, ,\nn\\
&&\p_i{^\a}=\cO_2\, ,\nn\\
&&\p_{ij}{^\a}=4\l_{ij}{}^{0\a}+\hcO\, .                    \lab{2.5}
\eea

In a similar manner, we can determine the asymptotic behavior of the
Hamiltonian multipliers (for instance, $\dot A^{ij}{_0}=u^{ij}{_0}$
implies $u^{ij}{_0}=\hcO$, etc.).

Now, we wish to check whether the global Poincar\'e generators
\eq{2.2} are well defined in the phase space characterized by the
asymptotic properties \eq{2.3}-\eq{2.5}.

\section{Improving the Poincar\'e generators}

In the Hamiltonian theory, the generators of symmetry
transformations act on dynamical variables via the Poisson bracket
operation, which is defined in terms of functional derivatives. A
functional
$$
F[\vphi,\p]=\int d^3x f(\vphi(x),\pd_\m\vphi(x),\p(x),\pd_\n\p(x))
$$
has well defined functional derivatives if its variation can be
written in the form
\be
\d F=\int d^3x\bigl[ A(x)\d\vphi(x)+B(x)\d\p(x)\bigr]\, ,   \lab{3.1}
\ee
where terms $\d\vphi_{,\m}$ and $\d\p_{,\n}$ are absent.

The global Poincar\'e generators \eq{2.2} do not satisfy these
requirements, as we shall see. This will lead us to improve their
form by adding certain surface terms, which turn out to define
energy-momentum and angular momentum of the gravitating system
\cite{3,18}.

\subsection{Spatial translation}

The variation of the spatial translation generator $P_\a$, given by
Eq. \eq{A4c}, yields
\bsubeq\lab{3.2}
\bea
\d P_\a=&&\int d^3x\d\cP_\a\, ,\nn\\
\d\cP_\a=&&\d\bcH_\a-\fr{1}{2}A^{ij}{_\a}\d\bcH_{ij}
 +2\l_{ij}{}^{0\b}\d\bcH^{ij}{}_{\a\b}+\p_k{^0}\d b^k{}_{0,\a}\nn\\
&&+\fr{1}{2}\p_{ij}{^0}\d A^{ij}{}_{0,\a}
  -\fr{1}{4}\l_{ij}{}^{\b\g}\d\p^{ij}{}_{\b\g,\a}
  -\fr{1}{2}\d(\l_{ij}{}^{\b\g}\p^{ij}{}_{\a\b})_{,\g}+R\,.\lab{3.2a}
\eea
Here, all terms that contain unwanted variations $\d\p_{,\a}$ or
$\d\vphi_{,\a}$ are written explicitly, while those that have the
correct, regular form \eq{3.1} are denoted by $R$. A simple formula
$\p_k{^0}\d b^k{}_{0,\a}=(\p_k{^0}\d b^k{}_0)_{,\a}+R$
allows us to conclude that $\p_k{^0}\d b^k{}_{0,\a}=\pd\hcO+R$, where
we used $\p_k{^0}=\hcO$, according to the asymptotic conditions
\eq{2.5}. Now, we apply the same reasoning to the terms proportional
to $\p_{ij}{^0}$, $\p^{ij}{}_{\a\b}$, $\p^{ij}{}_{0\b}$ (present in
$\bcH^{ij}{}_{0\b}$) and $A^{ij}{_\m}$, and find
\bea
\d\cP_\a&&=\d\bcH_\a +\pd\hcO +R\, .                       \lab{3.2b}
\eea
Using the explicit form of $\bcH_\a$, equation \eq{A1c}, we obtain
the result
\bea
\d\cP_\a&&=-\d(b^i{_\a}\p_i{^\g})_{,\g}+(\p_i{^\g}\d b^i{_\g})_{,\a}
           +\pd\hcO+R \nn\\
        &&=-\d(b^i{_\a}\p_i{^\g})_{,\g}+\pd\cO_3+R\, ,     \lab{3.2c}
\eea
\esubeq
where the last equality follows from $\p_i{^\g}\d b^i{_\g}=\cO_3$.
As a consequence, the variation of the spatial translation generator
$P_\a$ can be written in the simple form
\bea
&&\d P_\a=-\d E_\a +R\, ,\nn\\
&&E_\a\equiv\oint dS_\g (b^k{_\a}\p_k{^\g}) \, ,            \lab{3.3}
\eea
where $E_\a$ is defined as a surface integral over the boundary of
the three--dimensional space. This result allows us to redefine the
translation generator $P_\a$,
\be
P_\a\to \tP_\a=P_\a+E_\a\, ,                                \lab{3.4}
\ee
so that the new, improved expression $\tP_\a$ has {\it well defined
functional derivatives\/}.

The surface integral for $E_\a$ is {\it finite\/} since
$b^k{_\a}\p_k{^\g}=\cO_2$, in view of the asymptotic conditions
\eq{2.5}. While the old generator vanishes on shell (as an integral
of a linear combination of constraints), $\tP$ does not --- its
on-shell value is $E_\a$. Since $\tP_\a$ is the generator of the
asymptotic spatial translations, we expect that $E_\a$ will be the
value of the related conserved charge --- linear momentum; this will
be proved in Sec. 4.

\subsection{Time translation}

Similar procedure can be applied to the time translation generator
$P_0$:
\bsubeq\lab{3.5}
\bea
&&\d P_0=\int d^3x \d\cP_0\, ,\nn\\
&& \d\cP_0=\d\cH_T-\d(b^k{_0}\p_k{^\g})_{,\g} +\pd\hcO\, , \lab{3.5a}
\eea
where we used Eq. \eq{A4c} for $\cP_0$, and the adopted asymptotic
conditions for $A^{ij}{_0}$ and $\p^{ij}{}_{0\b}$. Since $\cH_T$
does not depend on the derivatives of momenta (on shell), as shown
in Appendix C of Ref. \cite{17}, we can write
\bea
\d\cH_T&&={\pd\cH_T\over\pd b^k{}_{\m,\a}}\d b^k{}_{\m,\a}+
{1\over 2}{\pd\cH_T\over\pd A^{ij}{}_{\m,\a}}\d A^{ij}{}_{\m,\a}+R\nn\\
&&\approx-{\pd\cL\over\pd b^k{}_{\m,\a}}\d b^k{}_{\m,\a}
-{1\over 2}{\pd\cL\over\pd A^{ij}{}_{\m,\a}}\d A^{ij}{}_{\m,\a}+R\, .\nn
\eea
The second term has the form $\pd\hcO+R$, so that
\be
\d\cH_T\approx -\pd_\a\left({\pd\cL\over\pd b^k{}_{\m,\a}}
      \d b^k{_\m}\right) +\pd\hcO+R =\pd\cO_3+R \, ,       \lab{3.5b}
\ee
\esubeq
and we find that
$$
\d\cP_0=-\d(b^k{_0}\p_k{^\g})_{,\g} +\pd\cO_3+R\, .
$$
Hence,
\bea
&&\d P_0= -\d E_0 + R\, , \nn\\
&&E_0=\oint dS_\g (b^k{_0}\p_k{^\g})\, .                    \lab{3.6}
\eea
The improved time translation generator
\be
\tP_0=P_0+E_0                                               \lab{3.7}
\ee
has well defined functional derivatives, and the surface term $E_0$
is finite on account on the adopted asymptotics. As we shall see,
the on-shell value $E_0$ of the time translation generator
represents the energy of the gravitating system.

Expressions \eq{3.6} and \eq{3.3} for the energy and momentum can be
written in a Lorentz covariant form:
\be
E_\m=\oint dS_\g (b^k{_\m}\p_k{^\g})\, .                    \lab{3.8}
\ee

It is interesting to observe that the value of the energy $E_0$ can
be calculated from the formula $E_0=\int d^3x\cH_T$. Indeed,
$$
\int d^3x \cH_T
  \approx \int d^3x\pd_\g\bar D^\g =\oint dS_\g\bar D^\g=E_0\, ,
$$
since $\bar D^\g=b^k{_0}\p_k{^\g}+\hcO$.

\subsection{Rotation}

Next, we want to check if the rotation generator $M_{\a\b}$ has well
defined functional derivatives:
\bea
&&\d M_{\a\b} = \int d^3x\,\d{\cM}_{\a\b}\,,  \nn\\
&&\d{\cM}_{\a\b} = x_{\a}\d{\cP}_{\b}-x_{\b}\d{\cP}_{\a}+
  \d\p_{\a\b}{}^{\g}{}_{,\,\g} + R \, ,                     \lab{3.9}
\eea
where $\p_{\m\n}{^\r}=\d^i_\m\d^j_\n\p_{ij}{^\r}$. Using the
expression \eq{3.2c} for $\d\cal P_{\a}$, one finds
\bea
&&\d M_{\a\b} = -\d E_{\a\b} + R \,, \nn\\
&&E_{\a\b}=\oint dS_{\g}\left[
  x_\a\bigl(b^k{_\b}\p_k{^\g}\bigr)-x_\b\bigl(b^k{_\a}\p_k{^\g}\bigr)
  -\p_{\a\b}{}^{\g}\right]\,.                              \lab{3.10}
\eea
In the course of obtaining the above expression, the term
$\oint dS_{[\a}\,x_{\b]}\,\phi \,$
($\phi\equiv \p_k{^\g}\d b^k{_\g}$) has been discarded as a
consequence of $dS_{\a}\propto x_{\a}$ on the integration sphere.
The corresponding improved rotation generator is given by
\be
\tM_{\a\b} \equiv M_{\a\b} + E_{\a\b}  \, .                \lab{3.11}
\ee

Although $\tM_{\a\b}$ has well defined functional derivatives, the
assumed asymptotics \eq{2.3}-\eq{2.5} does not ensure the finiteness
of $E_{\a\b}$ owing to the presence of $\cO_1$ terms. Note, however,
that the actual asymptotics is refined by the principle of
arbitrarily fast decrease of all on-shell vanishing expressions.
Thus, we can use the constraints $\bcH_\a$ and $\bcH_{ij}$, in the
lowest order in $r^{-1}$, to conclude that
\be
\p_{\a}{}^{\b}{}_{,\,\b} = \cO_4 \,,\qquad
2\p_{[\a\b]} + \p_{\a\b}{}^{\g}{}_{,\,\g} = \cO_3 \, ,     \lab{3.12}
\ee
where $\p_{\m\n}=\d^i_\m\eta_{\n\r}\p_i{^\r}$. As a consequence, the
angular momentum density decreases like $\cO_3\,$:
\be
\left(2x_{[\a}\p_{\b]}{}^{\g}
- \p_{\a\b}{}^{\g}\right)_{,\,\g} = \cO_3\, .              \lab{3.13}
\ee
As all variables in the theory are assumed to have asymptotically
polynomial behaviour in $r^{-1}$, the integrand of $E_{\a\b}$ must
essentially be of an $\cO_2$ type to agree with the above
constraint. The possible $\cO_1$ terms are divergenceless, and do
not contribute to the corresponding surface integral (as shown in
Appendix B). This ensures the finiteness of the rotation generator.

\subsection{Boost}

By varying the boost generator \eq{2.2b}, we find:
\bea
&&\d M_{0\b} = \int d^3x\,\d{\cM}_{0\b}\,,  \nn\\
&&\d{\cM}_{0\b} = x_0\d{\cP}_{\b}-x_{\b}\d{\cP}_0+
  \d\p_{0\b}{}^{\g}{}_{,\,\g} + R \, ,
                                                           \lab{3.14}
\eea
in analogy with \eq{3.9}. Here, we need to calculate $\d{\cal P}_0$,
equation \eq{3.5a}, up to terms $\pd\cO_4\,$. A simple calculation
gives
\be
\d{\cM}_{0\b} = \d\left( \p_{0\b}{}^{\g}
 -x_0 b^k{_{\b}}\p_k{^{\g}}+x_{\b}b^k{_0}\p_k{^{\g}}\right)_{,\,\g}
 +\left(x_{\b}X^{\g}\right)_{,\,\g} + \pd\cO_3 + R \,,     \lab{3.15}
\ee
where the term
\[
X^{\g} \equiv {{\pd\cL}\over {\pd b^k{_{\m ,\,\g}}}}\d b^k{_{\m}}
\]
is not a total variation, nor does it vanish on account of the
asymptotic conditions \eq{2.3}-\eq{2.5}. To get rid of this unwanted
term, we shall further restrict the phase space by adopting the
following {\it parity conditions\/}:
\bsubeq \lab{3.16}
\be
b_{i\m}=\eta_{i\m} + {{p_{i\m}(\mb{n})}\over r} +
        {{q_{i\m}(t,\mb{n})}\over {r^2}} +\cO_3 \,,       \lab{3.16a}
\ee
where $\mb{n} = \mb{x} /r\,$ is the three-dimensional unit vector,
and
\bea
&&p_{i\m}(\mb{n})=p_{i\m}(-\mb{n})\,, \nn\\
&&q_{i\m} =q_{i\m}^{(1)}(\mb{n}) +t q_{i\m}^{(2)}(\mb{n})\,,\quad
q_{i\m}^{(2)}(\mb{n}) = -q_{i\m}^{(2)}(-\mb{n})\,.        \lab{3.16b}
\eea
\esubeq
Time independence of $p_{i\m}$, and linear time dependence of
$q_{i\m}$ are a consequence of the required invariance of asymptotic
conditions under the global Poincar\'e transformations. It is
straightforward to verify that the {\it refined asymptotics ensures
the vanishing of the unwanted $X$ term in \eq{3.15}\/}. Therefore,
the improved boost generator, with well defined functional
derivatives, has the form
\bea
&&\tM_{0\b} \equiv M_{0\b} + E_{0\b} \,, \nn\\
&&E_{0\b}=\oint d S_{\g}\left[
  x_0\bigl(b^k{_\b}\p_k{^\g}\bigr)- x_\b\bigl(b^k{_0}\p_k{^\g}\bigr)
  - \p_{0\b}{}^{\g}\right]\,.                              \lab{3.17}
\eea

It remains to be shown that the adopted asymptotics ensures the
finiteness of $E_{0\b}$. That this is indeed true can be seen by
analyzing the constraints $\bcH_\ort$ and $\bcH_{ij}$ at spatial
infinity. Using the needed formulas from Appendix A, one finds:
\be
\p_0{^{\b}}{}_{,\,\b} = \cO_4 \,,\qquad
\p_{0\b} + \p_{0\b}{}^{\g}{}_{,\,\g} = \cO_3 \,.           \lab{3.18}
\ee
One should observe that this result holds in any teleparallel
theory; indeed, the relation $\hat\cH_T \equiv \cH_T-\pd_\a\bar D^\a
\approx 0$ implies $\p_0{^{\b}}{}_{,\,\b}\approx \cH_T +\cO_4\approx
\p_A\dot\vphi^A -\cL +\cO_4 = \cO_4\,$. It is now easy to verify
that the boost density decreases like $\cO_3$ for large $r$. Then,
the arguments given in Appendix B lead us to conclude that $E_{0\b}$
is finite.

The improved boost generator $\tM_{0\b}$ is a well defined functional
on the phase space defined by the refined asymptotic conditions
\eq{2.3}--\eq{2.5} and \eq{3.16}. The importance of suitably chosen
parity conditions for a proper treatment of the angular momentum has
been clearly recognized in the past \cite{3,21}.

Using the Lorentz 4-notation, we can write:
\bea
&&\tM_{\m\n} \equiv M_{\m\n} + E_{\m\n} \,, \nn\\
&&E_{\m\n}=\oint d S_\g\left[
            x_\m(b^k{_\n}\p_k{^\g})- x_\n(b^k{_\m}\p_k{^\g})
           -\p_{\m\n}{}^{\g}\right]\,.                     \lab{3.19}
\eea

\section{Conserved quantities}

In the preceding section, we obtained the improved Poincar\'e
generators ($\tP_\m\,$, $\tM_{\m\n}$). Their action on the fields
and momenta is the same as before, since surface terms act trivially
on local quantities. Explicit transformations are obtained by the
replacement \eq{2.1} in the local expressions of Ref. \cite{17}.
Once we know the generators $\tP_\m$ and $\tM_{\m\n}$ have the
standard Poincar\'e action on the whole phase space, we deduce their
algebra to be that of the Poincar\'e group:
\bea
&&\left\{\tP_\m \,,\tP_\n \right\} = 0  \, , \nn\\
&&\left\{\tP_\m \,,\tM_{\n\l} \right\} =
  \eta_{\m\n}\tP_\l - \eta_{\m\l}\tP_\n \, , \nn\\
&&\left\{\tM_{\m\n}\,,\tM_{\l\r} \right\}=
  \eta_{\m\r}\tM_{\n\l}- \eta_{\m\l}\tM_{\n\r}-(\m\lra\n)\,.\lab{4.1}
\eea
The line of reasoning that leads to the above result  does not
guarantee the strong equalities in \eq{4.1}. They are rather
equalities up to trivial generators, such as squares of constraints
and surface terms. In fact, the latter are not expected to appear in
\eq{4.1} as a consequence of the result of Ref. \cite{22} that
the Poisson bracket of two well defined generators is necessarily a
well defined generator. This is of particular importance for the
existence of conserved quantities in the theory. In what follows, we
shall explicitly verify the absence of these surface terms, and
prove the conservation of all the symmetry generators.

The result \eq{4.1} is an expression of the asymptotic Poincar\'e
invariance of the theory. In what follows, our task will be to show
that this symmetry implies, as usual, the existence of conserved
charges.

The general method for constructing the generators of local
symmetries in the Hamiltonian approach has been developed by
Castellani \cite{23}. A slight modification of that method
can be applied to study global symmetries. One can show that
necessary and sufficient conditions for a phase-space functional
$G[\vphi,\p,t]$ to be a generator of global symmetries take the
form:
\bsubeq\lab{4.2}
\bea
\left\{ G \,,\tH\right\}+{\pd G\over\pd t}\,&&=C_{PFC}\,,\lab{4.2a}\\
\left\{ G \,,\phi_m \right\}&&\approx 0 \,,              \lab{4.2b}
\eea
\esubeq
where $\tH$ is the improved Hamiltonian, $C_{PFC}$ is a primary
first class constraint, $\phi_m\approx 0$ are all the constraints in
the theory, and as before, the equality sign denotes an equality up
to surface terms and squares of constraints. The improved Poincar\'e
generators $\tP_\m\,$, ${\tilde M}_{\m\n}$ are easily seen to
satisfy \eq{4.2b}, as they are given, up to surface terms (whose
action on local quantities is trivial), by volume integrals of
first-class constraints. Having in mind that the improved
Hamiltonian $\tilde H$ equals $\tP_0$, one can verify that the
condition \eq{4.2a} is also satisfied, as a consequence of the part
of the algebra \eq{4.1} involving $\tP_0$. The condition \eq{4.2a}
is related to conservation laws; this is clearly seen if we rewrite
it as a weak equality
\be
{dG\over dt}\equiv \left\{ G \,,\tP_0\right\}+{\pd G\over\pd t}
            \approx S \,,                                   \lab{4.3}
\ee
where $S$ denotes possible surface terms. We see that the generator
$G$ is conserved only if these surface terms are absent. In what
follows, we shall explicitly evaluate $dG/dt$ for each of the
generators $G=\tP_\m,\tM_{\m\n}\,$, and check their conservation.

\subsub{1.} Let us begin with the energy. First, we note that
$\tP_0$, being a well defined functional, must commute with itself:
\[
\left\{ \tP_0 \,,\tP_0 \right\}=0 \, .
\]
Furthermore,
\[
{\pd\tP_0\over\pd t}={\pd H_T\over\pd t}=C_{PFC}\approx 0 \,,
\]
since the only explicit time dependence of the total Hamiltonian is
due to the arbitrary multipliers, which are always multiplied by
the primary first class constraints. Therefore,
\be
{d\tP_0\over dt}\approx {d E_0\over dt} \approx 0 \,,       \lab{4.4}
\ee
and we see that the surface term $E_0$, representing the
on-shell value of the energy, is a conserved quantity.

\subsub{2.} The linear momentum and the spatial angular momentum
have no explicit time dependence. To evaluate their Poisson brackets
with $\tP_0$, we shall use the following procedure. Our improved
generators are (non-local) functionals, having the form of integrals
of some local densities. The Poisson bracket of two such generators
can be calculated by acting with one of the generators on the
integrand of the other. In the case of linear momentum, we have
\[
\left\{ \tP_0 \,,\tP_\a \right\} = \int d^3 x\,
\left\{ \hat\cH_T + \pd_\g\p_0{^\g}\,, \tP_\a \right\}
\approx \int d^3 x\,\pd_\g\left\{ \p_0{^\g}\,, \tP_\a \right\}\, ,
\]
because $\hcH_T\approx 0$ is a constraint in the theory, and
therefore, weakly commutes with all the symmetry generators. The
last term in the above formula is easily evaluated (Appendix C),
with the final result
\be
\left\{ \tP_0 \,,\tP_\a \right\} \approx
\oint dS_\g \,\pd_\a\p_0{^\g} = 0 \,,                       \lab{4.5}
\ee
as a consequence of $\pd_\a\p_0{^\g}=\cO_3\,$. Therefore, no
surface term appears in \eq{4.3} for $G=\tP_\a$, and we have the
conservation law
\be
{{d\tP_\a}\over {dt}}\approx {{d E_\a}\over {dt}}
                       \approx 0 \,.                        \lab{4.6}
\ee

\subsub{3.} In a similar way, we can check the conservation of the
rotation generator. Using the results of Appendix C, we find:

\bea
\left\{ \tP_0 \,,{\tM}_{\a\b} \right\}&\approx &
\int d^3 x\,\pd_\g\left\{ \p_0{^\g}\,, {\tM}_{\a\b} \right\}
\approx\int d^3 x\,(x_\a\pd_\b - x_\b\pd_\a)\,\pd_\g \p_0{^\g} \nn\\
&\approx &\int d^3 x\,(\pd_\b x_\a - \pd_\a x_\b)\,\pd_\g \p_0{^\g}
\approx\oint (x_\a dS_\b - x_\b dS_\a)\,\pd_\g \p_0{^\g}=0\,, \nn
\eea
because $\pd_\g \p_0{^\g}=\cO_4$ according to \eq{3.18}. Therefore,
the surface term in \eq{4.3} is absent for $G=\tM_{\a\b}$, and the
on-shell value of the rotation generator is conserved:
\be
{{d \tM_{\a\b}}\over dt}\approx {d E_{\a\b}\over dt}
                        \approx 0\,.                        \lab{4.7}
\ee

\subsub{4.} Finally, the boost generator \eq{2.2b} has an explicit,
linear dependence on time, and satisfies
\be
\left\{ \tM_{0\b}\,,\tP_0 \right\}+{\pd\tM_{0\b}\over\pd t} =
\left\{ \tM_{0\b}\,,\tP_0 \right\}+\tP_\b \approx
 -\left\{ \tP_0\,,\tM_{0\b}\right\}+E_\b \, .               \lab{4.8}
\ee
The evaluation of the Poisson bracket in \eq{4.8} is done with the
help of Appendix C:
\be
\left\{ \tP_0\,,{\tM}_{0\b} \right\} \approx
\int d^3 x\,\pd_\g\left[ \p_{\b}{}^{\g} - \pd_\a
\left( x_\b {{\pd\cL}\over {\pd b^0{_{\a\,,\g}}}}\right)\right]=
\int d^3 x\,\pd_\g \p_{\b}{}^{\g} = E_\b \,,                \lab{4.9}
\ee
where we used the antisymmetry of $\pd\cL/\pd b^0{_{\a\,,\g}}$ in
$(\a\,,\g)$,  and $\pd_\g \p_0{^\g}=\cO_4$. As opposed
to all the other generators, the Poisson bracket of the boost
generator with $\tP_0$ does not vanish: its on-shell value is
precisely the value of the linear momentum $E_\b$. Substitution of
this result back into \eq{4.8} yields the boost conservation law
\cite{f2}:
\be
{d\tM_{0\b}\over dt}\approx{d E_{0\b}\over dt}\approx 0\, .\lab{4.10}
\ee

In conclusion, {\it all ten Poincar\'e generators of the general
teleparallel theory are conserved quantities\/} in the phase space
defined by the appropriate asymptotic and parity conditions.

\section{Lagrangian form of the conserved charges}

In this section,we wish to transform our Hamiltonian expressions for
the conserved quantities into the Lagrangian form, and compare the
obtained results with the related GR expressions. This will be
achieved by expressing all momentum variables, in $E_\m$ and
$E_{\m\n}$, in terms of the fields and their derivatives, using the
defining relations $\p_A=\pd\cL/\pd\dot\vphi^A$. All calculations
refer to the one-parameter teleparallel theory ($2A+B=1$, $C=-1$).

\subsection{Energy and momentum}

The energy--momentum expression \eq{3.8} can be transformed into the
Lagrangian form with the help of the relation \eq{D3} that defines
$\p_i{^\g}$:
\be
E_\m=\oint dS_\g H_\m{}^{0\g} \, ,\qquad
H_\m{}^{0\g}\equiv-4bb^i{_\m}h^{j\g}\b_{ij}{^0} \, .        \lab{5.1}
\ee
Now, we wish to compare this result with GR.

Using the decomposition of $f^i{_\m}\equiv b^i{_\m}-\d^i_\m$ into
symmetric and antisymmetric part, $f_{i\m}=s_{i\m}+a_{i\m}$, and the
asymptotic conditions \eq{2.3}--\eq{2.5}, we obtain
$$
H_0{}^{0\g}=2a(s_c{}^{c,\g}-s_c{}^{\g,c})-2aa^{c\g}{}_{,c}+\cO_3\, .
$$
Note that the second term in $\b_{ij}{^0}$, proportional to $(2B-1)$
and given by Eq. \eq{D4}, does not contribute to this result. Now,
after introducing the  well known superpotential of
Landau and Lifshitz,
$$
h^{\m\n\l}=\pd_\r\psi^{\m\n\l\r}\, ,\qquad
\psi^{\m\n\l\r}\equiv
          a(-g)(g^{\m\n}g^{\l\r}-g^{\m\l}g^{\n\r})\, ,
$$
one easily verifies that
$h^{00\g}=2a(s_c{}^{c,\g}-s_c{}^{\g,c})+\cO_3$. Then, after
discarding the inessential divergence of the antisymmetric tensor in
$H_0{}^{0\g}$, we find the following Lagrangian expression for
$E_0$:
\be
E_0=\oint dS_\g h^{00\g} \, .                               \lab{5.2}
\ee
Thus, the energy of the one-parameter teleparallel theory is given
by the {\it same\/} expression as in GR.

In a similar manner, we can transform the expression for $E_\a$.
Starting with equation \eq{D4}, we note that the first term in
$\b_{ij}{^0}$, which corresponds to \tgr\ ($B=1/2$), gives the
contribution
\bea
(H_\a{}^{0\g})_\one=&&2a\bigl[\eta^{\g\b}(s_{\a\b,0}
       -s_{0\a,\b})+\d_\a^\g(s^c{}_{0,c}-s^c{}_{c,0})\bigr]\nn\\
     &&+4a(\d^{[\g}_\a a^{\b]}{_0})_{,\b}+\cO_3 \nn \, ,
\eea
which, after dropping the irrelevant divergence of the antisymmetric
tensor, can be identified with $h_\a{}^{0\g}$. The contribution of
the second term has the form
\bea
(H_\a{}^{0\g})_\two\,&&=a(2B-1)bb^i{_\a}h^{j\g}h^{k0}\TA_{ijk}\nn\\
                       &&= a(2B-1)\TA_\a{}^{\g 0}+\cO_3\nn \, ,
\eea
where $\TA_{ijk}=T_{ijk}+T_{kij}+T_{jki}$. Thus, the complete linear
momentum takes the form
\be
E_\a=\eta_{\a\m}\oint dS_\g
\left[ h^{\m 0\g}-a(2B-1)\TA{}^{\m 0\g}\right] \, ,         \lab{5.3}
\ee
which is {\it different\/} from what we have in GR.

Energy and momentum expressions \eq{5.2} and \eq{5.3} can be written
in a Lorentz covariant form as
\bea
&&E^\m =\oint dS_\g \,\bar h^{\m 0\g}
       =\int d^3 x\, \bar \th^{\m 0}\, ,\nn\\
&&\bar\th^{\m\n}\equiv \bar h^{\m\n\r}{}_{,\r} \ ,\quad
 \bar h^{\m\n\r}\equiv h^{\m\n\r}-a(2B-1)\TA{}^{\m\n\r}\,.  \lab{5.4}
\eea
In the case $2B-1=0$, corresponding to \tgr\,, we see that
$\bar\th^{\m\n}$ coincides with the  Landau--Lifshitz
symmetric energy--momentum complex $\th^{\m\n}\equiv
h^{\m\n\r}{}_{,\r}$ of GR. When $2B-1\neq 0$, the momentum acquires
a {\it correction\/} proportional to the totally antisymmetric part
of the torsion.

\subsection{Angular momentum}

The elimination of momenta from Eq. \eq{3.19} leads to
\be
E_{\m\n}=\oint dS_\g\bigl( x_\m H_\n{}^{0\g}-x_\n H_\m{}^{0\g}
            -4\l_{\m\n}{}^{0\g}\bigr)\, ,                   \lab{5.5}
\ee where we used the expression \eq{D3} for $\p_k{^\a}$, and
$\pi_{\m\n}{^\g}\approx 4\d^i_\m\d^j_\n\l_{ij}{}^{0\g}
\equiv 4\l_{\m\n}{}^{0\g}$.

In order to compare this result with GR, we shall first
eliminate $\l$ using the second field equation \eq{1.3b}. This is
most easily done in the gauge $A^{ij}{}_{\m}=0$. In what follows,
all the calculations and results refer to this gauge. Thus, Eq.
\eq{1.3b} in the gauge fixed form gives:
\be
4\pd_\g\l_{ij}{}^{0\g}\approx 8b\b _{[ij]}{^0}+\s^0{}_{ij}\,.\lab{5.6}
\ee
The first term in $\b_{ij}{^0}$ is given in Eq. \eq{D4}, and
corresponds to \tgr\ $(2B-1=0)$; in particular,
$4b(\b_{[ij]}{^0})_\one\approx -a\pd_\g H_{ij}^{0\g}$. Then, we
define $\l_\one$ by ignoring the term $\s^0{}_{ij}$ in \eq{5.6}:
$$
\pd_\g\Bigl[
4(\l_{ij}{}^{0\g})_\one +2aH_{ij}^{0\g}\Bigr]\approx 0\, .
$$
The related contribution to the angular momentum has the form (see
Appendix D):
\bsubeq
\be
E^{\m\n}_\one=\oint dS_\g\bigl[ x^\m h^{\n 0\g}-x^\n h^{\m 0\g}
         +\psi^{\m 0\g\n} +\cO(f^2)\bigr] \, ,             \lab{5.7a}
\ee
where $\psi^{\m 0\g\n}$ satisfies the relation
$\pd_\g\psi^{\m 0\g\n}=h^{\m 0\n}-h^{\n 0\m}$, and $\cO(f^2)$
denotes terms quadratic in $f^i{_\m}$ and/or its derivatives.
The parity conditions \eq{3.16} ensure that the $\cO(f^2)$ terms do
not contribute to the surface integral, hence
\be
E^{\m\n}_\one=\oint dS_\g K^{\m\n 0\g}
         =\int d^3 x (x^\m\th^{\n 0}-x^\n\th^{\m 0})\,.    \lab{5.7b}
\ee
\esubeq
Here, the tensor
$K^{\m\n\l\r}= x^\m h^{\n\l\r}-x^\n h^{\m\l\r}+\psi^{\l\m\n\r}$
satisfies the relation
$$
\pd_\r K^{\m\n\l\r}=x^\m\th^{\n\l}-x^\n\th^{\m\l} \, ,
$$
and $E^{\m\n}_\one$ represents the angular momentum of GR, as
expected.

The second term of $\b_{ij}{^0}$ in Eq. \eq{D4} defines the second
term of $\l=\l_\one+\l_\two$:
$$
4\pd_\g(\l_{ij}{}^{0\g})_\two
   \approx-2a(2B-1)bh^{k0}\TA_{ijk}+\s^0{}_{ij}\, .
$$
The related angular momentum contribution takes the form
\be
E_\two^{\m\n}= \oint dS_\g\Bigl[ a(2B-1)
  \bigl(x^\m\TA{}^{\n\g 0}-x^\n\TA{}^{\m\g 0}\bigr)
 -4\eta^{i\m}\eta^{j\n}(\l_{ij}{}^{0\g})_\two \Bigr] \, ,   \lab{5.8}
\ee
where the inessential $\cO(f^2)$ terms are ignored.

The complete angular momentum can be transformed into the form of a
volume integral:
\bea
E^{\m\n}=\int d^3 x \Bigl\{
 &&\bigl(x^\m\bar\th^{\n 0}-x^\n\bar\th^{\m 0}\bigr)
   +2a(2B-1) \TA{}^{\n\m 0} \nn\\
 &&+\eta^{i\m}\eta^{j\n}\Bigl[
    2a(2B-1)bh^{k0}\TA_{ijk}-\s^0{}_{ij}\Bigr] \Bigr\}\, .  \lab{5.9}
\eea
Note that the integrand of $E^{\m\n}$ is not of the simple form
$x^\m\bar\th^{\n 0}-x^\n\bar\th^{\m 0}$, since the corrected
energy-momentum complex $\bar\th^{\m\n}$ is (in general) not
symmetric.  We see that even in \tgr\ ($2B-1=0$), the
angular momentum differs from the GR expression (by the contribution
of the $\s^0{}_{ij}$ term).

 Our results for the conserved charges are valid for any
solution satisfying the asymptotic conditions defined by equations
\eq{2.3}--\eq{2.5} and \eq{3.16}. Assuming that the gravitational
field is produced by the Dirac field as a source, Hayashi and
Shirafuji \cite{9} found a specific solution for which
$f_{[0\a]}=k\ve_{\a\b\g}{n^\b S^\g/r^2}$, where $S^\g$ is the spin
of the source, and $k$ is a constant. The solution is obtained in
the weak field approximation, and yields a non-vanishing
antisymmetric part of the torsion. Using our expressions for the
conserved charges, we verified that this solution gives a vanishing
correction to the GR results. This is in agreement with the results
of Kawai and Toma \cite{24}, who studied the Noether charges in a
non-standard formulation of the teleparallel theory, and concluded
that all contributions stemming from $\TA_{ijk}$ effectively vanish.

\section{Concluding remarks}

In this paper, we presented an investigation of the connection
between the asymptotic Poincar\'e symmetry of spacetime and the
related conservation laws of energy-momentum and angular momentum in
the general teleparallel theory of gravity.

The generators of the global Poincar\'e symmetry in the asymptotic
region are derived from the related gauge generators, constructed in
Ref. \cite{17}. Since these generators act on dynamical variables
via the Poisson bracket operation, it is natural to demand that they
have well defined functional derivatives in a properly defined phase
space. This requirement leads to the conclusion that the Poincar\'e
generators have to be improved by adding certain surface terms,
which represent the values of energy-momentum and angular momentum
of the gravitating system.

The general asymptotic behavior of dynamical variables, determined
by Eqs. \eq{2.3}--\eq{2.5}, is sufficient to guarantee the existence
of well behaved (finite and differentiable) generators of spacetime
translations and spatial rotations. This is, however, not true for
boosts: they can be improved by adding surface integrals only if one
imposes the additional parity conditions \eq{3.16}. Using the
canonical criterion \eq{4.2}, we were able to show that the improved
generators are not only finite, but also conserved; hence, the
related charges, energy-momentum and angular momentum, are the
constants of motion.

Our results for energy-momentum and angular momentum are valid for
the {\it general teleparallel theory\/}.  In the context of \tgr,
Nester \cite{25} used some geometric arguments to derive an
energy-momentum expression, which agrees with our formula \eq{3.8}.
On the basis of this result, he was able to formulate a pure
tensorial proof of the positivity of energy in GR, in terms of the
teleparallel geometry. Using the constraint
$\p_{\ort\bk}+2aJT^\bm{}_{\bm\bk}\approx 0$, which holds in \tgr\
(Appendix A), one can derive another equivalent form of the energy
integral, $E_0=-2a\int d^3x\pd_\g T_a{}^{a\g}$, appearing in the
literature \cite{26}.

After transforming the obtained surface integrals to the Lagrangian
form, one finds that the conserved charges in the one-parameter
teleparallel theory {\it differ\/} (in general) from the
corresponding GR expressions. Mielke and Wallner \cite{x11} discussed
Lagrangian form of the energy-momentum of some exact solutions in
the teleparallel limit of PGT. A purely Lagrangian analysis of the
energy-momentum in the specific case of \tgr\ without matter has
been carried out in Ref. \cite{27}, with the same result as in GR.
Kawai and Toma \cite{24} studied Noether charges for both
energy-momentum and angular momentum, in a non-standard formulation
of the teleparallel theory, and found that these charges have the
{\it same\/} form as in GR. Such a conclusion can be understood by
noting that, effectively, they used specific boundary conditions
corresponding to the Hayashi--Shirafuji solution \cite{9}, for which
the conserved charges are indeed of the same form as in GR.

The results obtained in this paper can be used to justify some
proposals used in the literature for the energy-momentum in \tgr\
\cite{24,25,26,27}, extend the concepts of energy-momentum and
angular momentum from \tgr\ to the general teleparallel theory, and
study the related stability properties.

\section*{Acknowledgments}

This work was partially supported by the Serbian Science Foundation,
Yugoslavia. One of us (M. B.) would like to acknowledge the support
of the Science Foundation of Slovenia.

\appendix

\section{Hamiltonian and gauge generators}

The canonical dynamics of the general teleparallel theory \eq{1.2}
is described by the total Hamiltonian \cite{16}
\bsubeq\lab{A1}
\bea
&&\cH_T=\hat\cH_T+\pd_\a\bar D^\a\, ,\nn\\
&&\hat\cH_T\equiv \bcH_c+u^i{_0}\p_i{^0}+\fr{1}{2}u^{ij}{_0}\p_{ij}{^0}
 +\fr{1}{4}u_{ij}{}^{\a\b}\p^{ij}{}_{\a\b}+(u\cdot\phi)\, ,\nn\\
&&\bar D^\a=b^k{_0}\p_k{^\a}+\fr{1}{2}A^{ij}{_0}\p_{ij}{^\a}
  -\fr{1}{2}\l_{ij}{}^{\a\b}\p^{ij}{}_{0\b}\, ,             \lab{A1a}
\eea
where $\p_i{^0}$, $\p_{ij}{^0}$, $\p^{ij}{}_{\a\b}$ and $\phi$ are
primary constraints, $u$ are the corresponding multipliers, and
$\bcH_c$ is the canonical Hamiltonian,
\be
\bcH_c= N\bcH_\ort+N^\a\bcH_\a-\fr{1}{2}A^{ij}{_0}\bcH_{ij}
          -\l_{ij}{}^{\a\b}\bcH^{ij}{}_{\a\b} \, ,          \lab{A1b}
\ee
whose components are given by
\bea
&&\bcH_\a=\pi_i{^\b}T^i{_{\a\b}}-b^k{_\a}\nabla_\b\pi_k{^\b}
     +\fr{1}{2}\pi^{ij}{}_{0\a}\nabla_\b\l_{ij}{}^{0\b}\, ,\nn\\
&&\bcH_{ij}=2\pi_{[i}{^\a}b_{j]\a}+\nabla_\a\pi_{ij}{^\a}
     +2\pi^s{}_{[i0\a}\l_{sj]}{}^{0\a}\, ,\nn\\
&&\bcH^{ij}{}_{\a\b}=R^{ij}{}_{\a\b}
     -\fr{1}{2}\nabla_{[\a}\pi^{ij}{}_{0\b]}\, ,  \nn\\
&&\bcH_\ort=\cH_\ort -\fr{1}{8}(\pd\cH_\ort/\pd A^{ij}{}_\a)
                      \pi^{ij}{}_{0\a}\, ,\nn\\
&&\cH_\ort\equiv \hp_i{^\bk}T^i{}_{\ort\bk}-J\cL_T
                 -n^k\nabla_\b\p_k{^\b}\, .                 \lab{A1c}
\eea
\esubeq
Here, $\nabla_\m$ is the covariant derivative,
$n_k=h_k{^0}/\sqrt{g^{00}}$ is the unit normal to the hypersurface
$x^0=$ const, the bar over the Latin index is defined by the
decomposition $V_k=V_\ort n_k+V_\bk$, $V_\ort=n^kV_k$, of an
arbitrary vector $V_k$, the lapse and shift functions $N$ and $N^\a$
are given as $N=n_kb^k{_0}$, $N^\a=h_\bk{^\a}b^k{_0}$, $J$ is
determined by $b=NJ$, and $\hp_{i\bk}=\p_i{^\a}b_{k\a}$. Note that
$\bcH_c$ is linear in unphysical variables
$(b^k{_0},A^{ij}{_0},\l_{ij}{}^{\a\b})$.

\begin{center}
\doublerulesep .5pt
\begin{tabular}{l l l}
\multicolumn{3}{l}{Table 1.}                   \\ \hline\hline
\rule{0pt}{12pt} &~first class   &~second class\\ \hline
\rule[-1pt]{0pt}{15pt}
primary &~$\p_i{^0},\p_{ij}{^0},\p^{ij}{}_{\a\b}$
        &~$\phi_{ij}{^\a},\p^{ij}{}_{0\b}$     \\ \hline
\rule[-1pt]{0pt}{15pt}
secondary &~$\bcH_\ort,\bcH_\a,\bcH_{ij},\bcH^{ij}{}_{\a\b}$
          &~                                   \\ \hline\hline
\end{tabular}
\end{center}
The dynamical classification of the sure constraints is given
in Table 1, where
\be
\phi_{ij}{^\a}=\p_{ij}{^\a}-4\l_{ij}{}^{0\a}\, .             \lab{A2}
\ee

The only terms in the total Hamiltonian $\cH_T$ that depend on the
specific form of the Lagrangian are dynamical Hamiltonian $\cH_\ort$
and extra primary constraints $\phi$. We display here, for
completeness, the explicit form of these quantities for the
teleparallel formulation of GR (\tgr) \cite{16}:
\bsubeq\lab{A3}
\bea
&&\cH_\ort=\fr{1}{2}P_T^2-J\cL_T(\bT)-n^k\nabla_\b\p_k{^\b}\, ,\nn\\
&&P_T^2={1\over 2aJ}\left( \hp_{(\bi\bk)}\hp^{(\bi\bk)}
        -{1\over 2}\hp^\bm{_\bm}\hp^\bn{_\bn} \right)\, ,\nn\\
&&\cL_T(\bT)=a\left(\fr{1}{4}T_{m\bn\bk}T^{m\bn\bk}
        +\fr{1}{2}T_{\bm\bn\bk}T^{\bn\bm\bk}
        -T^\bm{}_{\bm\bk}T_\bn{}^{\bn\bk} \right) \, ,      \lab{A3a}
\eea
and
\bea
&&(u\cdot\phi)=\fr{1}{2}u^{ik}\bphi_{ik}\, ,\nn\\
&&\bphi_{ik}=\phi_{ik}-\fr{1}{4}a\bigl(
  \pi_i{^s}{}_{0\a}B_{sk}^{0\a}
  +\pi_k{^s}{}_{0\a}B_{is}^{0\a}\bigr)   \, ,\nn\\
&&\phi_{ik}=\hp_{i\bk}-\hp_{k\bi}+a\nabla_\a B^{0\a}_{ik}\, ,\lab{A3b}
\eea
\esubeq
where $B^{0\a}_{ik}\equiv\ve^{0\a\b\g}\ve_{ikmn}b^m_\b b^n_\g$.

The Poincar\'e gauge generator of the general teleparallel theory
\eq{1.2} has the form \cite{17}
\bsubeq\lab{A4}
\be
G=G(\o)+G(\xi)\, ,                                          \lab{A4a}
\ee
where
\bea
G(\o)=&&-\fr{1}{2}\dot\o^{ij}\p_{ij}{^0}-\fr{1}{2}\o^{ij}S_{ij}\, ,\nn\\
G(\xi)=&&\,-\dot\xi^0\bigl(
   b^k{_0}\p_k{^0}+\fr{1}{2}A^{ij}{_0}\p_{ij}{^0}
  +\fr{1}{4}\l_{ij}{}^{\a\b}\p^{ij}{}_{\a\b} \bigr)-\xi^0\cP_0 \nn\\
&&-\dot\xi^\a\bigl(
   b^k{_\a}\p_k{^0}+\fr{1}{2}A^{ij}{_\a}\p_{ij}{^0}
  -\fr{1}{2}\l_{ij}{}^{0\b}\p^{ij}{}_{\a\b}\bigr)-\xi^\a \cP_\a\, .
                                                            \lab{A4b}
\eea
In the above expressions, we used the following notation:
\bea
S_{ij}=&&-\bcH_{ij}+ 2b_{[i0}\p_{j]}{^0}
 +2A^s{}_{[i0}\p_{sj]}{^0}+2\l_{s[i}{}^{\a\b}\p^s{}_{j]\a\b}\, ,\nn\\
\cP_0\equiv &&\hat\cH_T=\cH_T-\pd_\a\bar D^\a\, ,\nn\\
\cP_\a= &&\bcH_\a-\fr{1}{2}A^{ij}{_\a}\bcH_{ij}
 +2\l_{ij}{}^{0\b}\bcH^{ij}{}_{\a\b} +\p_k{^0}\pd_\a b^k{_0}
 +\fr{1}{2}\p_{ij}{^0}\pd_\a A^{ij}{_0}\nn\\
        &&-\fr{1}{4}\l_{ij}{}^{\b\g}\pd_\a\p^{ij}{}_{\b\g}
 -\fr{1}{2}\pd_\g\bigl(\l_{ij}{}^{\b\g}\p^{ij}{}_{\a\b}\bigr)\, .
                                                            \lab{A4c}
\eea
\esubeq

\section{On surface terms}

In this Appendix, we shall discuss asymptotic properties of vector
fields, which are important for understanding the structure of
surface terms.

Consider a vector field \mb{A}, with the following asymptotic
behavior:
\be
\mb{A} = {\mb{a}\over r} + \cO_2 \ , \qquad
         \hbox{div}\, \mb{A} = \cO_3 \, ,                    \lab{B1}
\ee
where $\mb{a}=\mb{a}(\mb{n})$, $\mb{n}=\mb{x}/r$. We shall prove
that under these assumptions
\be
\oint_{S_{\infty}} \mb{A} \cdot d\mb{S} = \hbox{finite}\,,   \lab{B2}
\ee
where $S_\infty$ is the sphere at spatial infinity.

Relations \eq{B1} imply $\hbox{div}\,({\mb a}/r) = 0$, for all $r\ne
0$. Integrating $\hbox{div}(\mb{a}/r)$ over the region $V$ outside
the sphere $S_R$ of radius $R$, and using  Gauss
divergence theorem yields
$$
\oint_{S_{\infty}}{\mb{a}\over r}\cdot d\mb{S} -
\oint_{S_R}{\mb{a}\over r}\cdot d\mb{S}=0\,,\qquad \forall R\ne 0\,.
$$
Here, the integral over $S_\infty$ is independent of $R$, while the
integral over $S_R$ is linear in $R$:
$$
\oint_{S_R} {\mb{a}\over r}\cdot d\mb{S} =
R\oint \mb{a}\cdot\mb{n}\ d\O \, .
$$
Hence, we must have $\oint\mb{a}\cdot\mb{n}\ d\O =0$, which is
equivalent to
$$
\oint_{S_{\infty}} {\mb{a}\over r}\cdot d\mb{S} =0 \,.
$$
Now, using the first relation in \eq{B1} we easily verify the
statement \eq{B2}.

Note that the same line of reasoning in the case of
\be
\mb{B}= {\mb{b}\over r^2} + \cO_3 \ , \qquad
\hbox{div}\, \mb{B} = \cO_4\, ,                              \lab{B3}
\ee
{\it does not lead\/} to $\oint\mb{B}\cdot d\mb{S}=0$, as one might
naively expect. For example, the everywhere regular vector field
$$
\mb{B}={1+\sqrt{1+r^2}\over(1+r^2)^2}\,\mb{x}
$$
satisfies \eq{B3}, but  yields $\oint\mb{B}\cdot d\mb{S}=4\p$.

\section{Conservation laws}

To verify the conservation of the improved Poincar\'e generators, we
need their Poisson brackets with $\tP_0$. The essential part of these
brackets is the expression $\left\{ \p_0{^\g}\,, G \right\}$,
representing the action of the Poincar\'e generators on the local
quantity $\p_0{^\g}$. Using the known Poincar\'e transformation law
for the momenta, as given by the equation (5.2) of Ref. \cite{17}, we
can write:
\be
\d_0 \p_k{^\g} = \left\{ \p_k{^\g}\,, G \right\}\approx
\ve_k{^s}\p_s{^\g} + \ve^\g{_\b}\p_k{^\b} +
\ve^0{_\b}{{\pd\cL}\over {\pd b^k{_{\g ,\,\b}}}} -
(\ve^\m{_\n}x^\n +\ve^\m)\pd_\m \p_k{^\g} \,,                \lab{C1}
\ee
wherefrom we read the corresponding Poisson brackets. In the case of
$\p_0{^\g}$, we have:
\bea
&&\left\{\p_0{^\g}\,,\tP_\m\right\}\approx\pd_\m\p_0{^\g}\,,\nn\\
&&\left\{\p_0{^\g}\,,\tM_{\a\b}\right\}\approx
  \left(\d_{\a}^{\g}\p_{0\b} + x_\a \pd_\b \p_0{^\g}\right)-
  \left(\a \leftrightarrow \b \right) \,, \nn\\
&&\left\{\p_0{^\g}\,,\tM_{0\b}\right\}\approx \p_\b{^\g}+
  \eta_{\a\b}{{\pd\cL}\over{\pd b^0{_{\g ,\,\a}}}}+
  x_0\pd_\b \p_0{^\g}-x_\b \pd_0 \p_0{^\g} \,.               \lab{C2}
\eea
The last of the above equations can further be simplified by
using the Lagrangian equations of motion. Thus,
\[
\pd_0\p_0{^\g}\approx \pd_0{\pd\cL\over \pd b^0{_{\g ,\,0}}}
 \approx \pd_\a{\pd\cL\over \pd b^0{_{\a ,\,\g}}}
         +{\pd\cL\over\pd b^0{_{\g}}}\, ,
\]
so that the $x_\b\pd_0\p_0{^\g}$ part becomes
\be
x_\b\pd_0\p_0{^\g} =
  x_\b\pd_\a{\pd\cL\over \pd b^0{_{\a ,\,\g}}} + \cO_3 =
  \pd_\a \left(x_\b{\pd\cL\over\pd b^0{_{\a ,\,\g}}}\right)+
  \eta_{\a\b}{\pd\cL\over\pd b^0{_{\g ,\,\a}}}+\cO_3 \,.     \lab{C3}
\ee
Its substitution back into \eq{C2} gives the result used in
equation \eq{4.9}.

\section{Connection with the Lagrangian formalism}

In this Appendix we collect some formulas which simplify the
derivation of the Lagrangian form of the conserved charges.

{\bf 1.} We begin with the relation
\bsubeq\lab{D1}
\bea
&&\b_{ijk}=\fr{1}{2}a(\t_{jki}-\t_{ijk}+\t_{kij})
           -\fr{1}{4}a(2B-1)\TA_{ijk} \, , \nn\\
&&\t_{kij}=\eta_{k[i}T_{j]}-\fr{1}{2}T_{kij}\, .            \lab{D1a}
\eea
After using the identity
$2b\t_{kij}= -b_{k\l}\nabla_\r H_{ij}^{\l\r}$, where
$H_{ij}^{\m\n}\equiv b(h_i{^\m}h_j{^\n}-h_j{^\m}h_i{^\n})$,
one easily obtains
\bea
&&4b\b_{(ij)}{^\m}=ah^{k\m}b_{i\l}\nabla_\r H_{jk}^{\l\r}
                   +(i\lra j)\, , \nn\\
&&4b\b_{[ij]}{^\m}=-a\nabla_\r H_{ij}^{\m\r}
                   -a(2B-1)bh^{k\m}\TA_{ijk} \, .          \lab{D1b}
\eea
\esubeq
When we combine the last equation with the weak relation
$\nabla_\m(4b\b_{[ij]}{^\m}+\s^\m{}_{ij}/2)\approx 0$, which follows
from the equations of motion \eq{1.3}, we obtain the relation
\bsubeq\lab{D2}
\be
\nabla_\m\Bigl[ 2a(2B-1)bh^{k\m}\TA_{ijk}
    -\s^\m{}_{ij}\Bigr] \approx 0\, ,                       \lab{D2a}
\ee
which, in the weak field approximation, reads
\be
\pd_\r\Bigl[2a(2B-1)\TA{}_{\m\n}{^\r}-\s^\r{}_{\m\n}\Bigr]
                      \approx \cO(f^2)+\hcO=\cO_4\, ,       \lab{D2b}
\ee
\esubeq
where $\cO(f^2)$ denotes terms quadratic in $f^i{_\m}$ and/or its
derivatives.

{\bf 2.} The momentum $\p_i{^\g}$ is defined by the relation
\be
\p_i{^\g}=-4bh^{j\g}\b_{ij}{^0}\, .                          \lab{D3}
\ee
In order to simplify the calculations, we rewrite $\b_{ij}{^0}$ as
a sum of two terms:
\bea
&&\b_{ij}{^0} =(\b_{ij}{^0}) _\one +(\b_{ij}{^0}) _\two\, ,\nn\\
&&4b(\b_{ij}{^0})_\one= ah^{k0}(b_{i\l}\nabla_\r H_{jk}^{\l\r}
   +b_{j\l}\nabla_\r H_{ik}^{\l\r})
   -a\nabla_\r H_{ij}^{0\r} \, , \nn\\
&&4b(\b_{ij}{^0})_\two=-a(2B-1)bh^{k0}\TA_{ijk}\, .          \lab{D4}
\eea
The first term corresponds to \tgr\ $(2B-1=0)$, and its contribution
to the angular momentum $E_{\m\n}=\int d^3x\cE_{\m\n}$ has the form
\bsubeq\lab{D5}
\be
\cE_{\m\n}^\one=\pd_\g\left[ a\d^i_\m\d^j_\n H_{ij}^{0\g}
  -4bx_\m b^i{_\n}h^{j\g}(\b_{ij}{^0})_\one \right]
  -(\m\lra\n) +\hcO \, .                                    \lab{D5a}
\ee
Going now back to the surface integral, we can use the parity
conditions \eq{3.16} to conclude that the terms quadratic in
$f^i{_\m}$ and of order $r^{-2}$ are even, so that their
contribution to the surface integral vanishes. Hence, we continue by
keeping only linear terms:
\bea
\cE^{\m\n}_\one=&&a\pd_\g\left[ H^{\m\n 0\g}+x^\m\bigl( H^{\n\g 0\r}
  -H^{\g 0\n\r}-H^{\n 0\g\r}\bigr)_{,\r}\right]\nn\\
  &&-(\m\lra\n)+\pd\cO(f^2) +\hcO\, ,                       \lab{D5b}
\eea
\esubeq
where $H^{ij\m\n}=\eta^{im}\eta^{nj}H_{mn}^{\m\n}$. Using the
decomposition $H^{ij\m\n}=H_S^{ij\m\n}+H_A^{ij\m\n}$, where
$H_S^{ij\m\n}$ is symmetric and $H_A^{ij\m\n}$ antisymmetric under
the exchange of the pairs of indices $(ij)\lra(\m\n)$, and the
identities
\bea
\psi^{\m\n\l\r}&&=abh^{i\n}h^{j\l}H_{ij}^{\m\r}
 =2a\bigl(H_S^{\m\r\n\l}-\eta^{[\m\n}\eta^{\r]\l}\bigr)
                        +\cO(f^2)\,,            \nn\\
\psi^{ij\m\n}&&-\psi^{\m ji\n}=\psi^{ij\n\m}\, ,\nn
\eea
we can derive the relation
$$
a(H_S^{\n\g 0\r}-H_S^{\g 0\n\r}-H_S^{\n 0\g\r})_{,\r}=
  \psi^{\g\r\n 0}{}_{,\r}=h^{\n 0\g}\, .
$$
After substituting $H\to H_S$ and $H\to H_A$, respectively, in
Eq. \eq{D5b}, we find
\bea
&&\cE^{\m\n}_\one(H_S)=\pd_\g \bigl(\psi^{\m 0\g\n}
 +x^\m h^{\n 0\g}-x^\n h^{\m 0\g}\bigr)+\pd\cO(f^2)+\hcO\, ,\nn\\
&&\cE^{\m\n}_\one(H_A)=\pd_\g\pd_\b \left[ ax^\m\bigl(H_A^{0\n\g\b}
 +2H_A^{0[\g\n\b]}\bigr)-(\m\lra\n)\right]
 +\pd\cO(f^2)+\hcO\ \, .                                     \lab{D6}
\eea
The term $\cE^{\m\n}_\one(H_A)$ effectively vanishes, since the
expression in square brackets is antisymmetric in $(\g,\b)$, and
the final expression for the angular momentum $E^{\m\n}_\one$ has
the form \eq{5.7a}.


\begin{thebibliography}{99}
\bibitem{f1} The weights of irreducible torsion components in the
action, $\a,\b,\g$, are defined in terms of the parameters $A,B,C$
by the relations: $2A+B+C=\a+\b$, $2(2A+B)=3\a$, $2(2A-B)=\a-4\g/9$
(see the first reference in \cite{6}).
\bibitem{f2} In view of the present results, we believe that our
   earlier conclusion in Ref. \cite{18}, namely that the boost generator
   in the generic $R+T^2+R^2$ theory with massive tordions is not
   conserved (except in the rest frame), should be revised.
\end{thebibliography}
\end{document}